# Fractional Josephson Effect: A Missing Step Is A Key Step

**Physicists are searching for superconducting materials that can host Majoranas. New evidence for these elusive particles is provided by missing Shapiro steps in a Josephson effect mediated by an accidental Dirac semimetal.**

**Fan Zhang[1]\* and Wei Pan[2]\***

[1]*Department of Physics, University of Texas at Dallas, Richardson, TX, USA.*
[2]*Sandia National Laboratories, Albuquerque, NM and Livermore, CA, USA.*
*\*e-mail: zhang@utdallas.edu; wpan@sandia.gov.*

In a uniform superconductor, electrons form Cooper pairs that pick up the same quantum mechanical phase for their bosonic wavefunctions. This spontaneously breaks the gauge symmetry of electromagnetism. In 1962 Josephson predicted [1], and it was subsequently observed, that Cooper pairs can quantum mechanically tunnel between two weakly coupled superconductors that have a phase difference $\phi$. The resulting supercurrent is a *$2\pi$* periodic function of the phase difference $\phi$ across the junction. This is the celebrated Josephson effect. More recently, a fractional Josephson effect related to the presence of Majorana bound states – Majoranas – has been predicted for topological superconductors. This fractional Josephson effect has a characteristic *$4\pi$* periodic current-phase relation. Now, writing in *Nature Materials*, Chuan Li and colleagues [2] report experiments that utilize nanoscale phase-sensitive junction technology to induce superconductivity in a fine-tuned Dirac semimetal $Bi_{0.97}Sb_{0.03}$ and discover a significant contribution of *$4\pi$* periodic supercurrent in Nb-$Bi_{0.97}Sb_{0.03}$-Nb Josephson junctions under radio frequency irradiation.

Strong motivation for this work is provided by the search for, and manipulation of, Majoranas that have been predicted to exist in topological superconductors [3-7]. Such a Majorana must be localized at a topological defect, such as a vortex core or natural crystal termination, and pinned to the middle of the superconducting gap, which is the reference of zero energy. These hallmarks and the emergence of Majoranas are completely determined by the symmetry and topology of quasiparticles inside the superconductor. Amazingly, exciting Majoranas costs no energy yielding ground-state degeneracy, and adiabatically exchanging defects binding Majoranas noncommutatively transforms the system from one ground state to another manifesting non-abelian braiding statistics [6]. Quantum information encoded in this ground-state space can be manipulated via braiding operations that are protected from the environment. As such, Majorana-based topological quantum computing holds considerable promise, as one can embed quantum information in a non-local and decoherence-free fashion [3].

The experimental realization of a topological superconductor [3] is challenging; here is where Josephson junctions enter. It was suggested [4] that a behavior analogous to a topological superconductor could be obtained by coupling the helical surface state of a topological insulator with a conventional *s*-wave Josephson junction (Fig. 1a). Across the junction there is "a particle in a box", except that the "potentials of the box" are provided by the superconductors. It follows that discrete Andreev bound states emerge at the junction and their spectrum is a function of the phase difference $\phi$ across the junction (Fig. 1b). It turns out that the bound states at *E = 0* are

Majoranas. Since $I \sim dE/d\phi$, the presence of Majoranas gives rise to a *4π periodic* current-phase relation (green lines in Fig. 1c). Under a dc bias voltage $V_{dc}$ across the junction, while charge-*2e* Cooper pairs tunnel at a frequency of *2eV$_{dc}$/h*, Majorana-mediated charge-*e* quasiparticles tunnel at half the frequency, *eV$_{dc}$/h* [7].

Experimentally observing this *4π periodic* fractional Josephson effect is complicated by the existence of multiple channels along the junction and by scattering from thermally excited bulk quasiparticles [7]. Both relaxation processes might gap out the Majoranas (red dashed lines in Fig 1b), lift the ground-state degeneracy, and thus restore the *2π periodic* supercurrent (red dashed lines in Fig. 1c). The fast high-frequency experiment performed by Li and colleagues [2] has the advantage of avoiding these relaxation processes. When a Josephson junction is irradiated with microwaves the dc current-voltage characteristic exhibits a Shapiro step [8] when the frequency of tunneling current is an integer multiple of the microwave frequency (Fig. 1d). All integer Shapiro steps should be present for an usual Josephson effect, whereas only the even steps should be visible for a *4π periodic* fractional Josephson effect [9]. Now, Li and colleagues [2] clearly find that the $N = \pm 1$ steps are missing in their experiment.

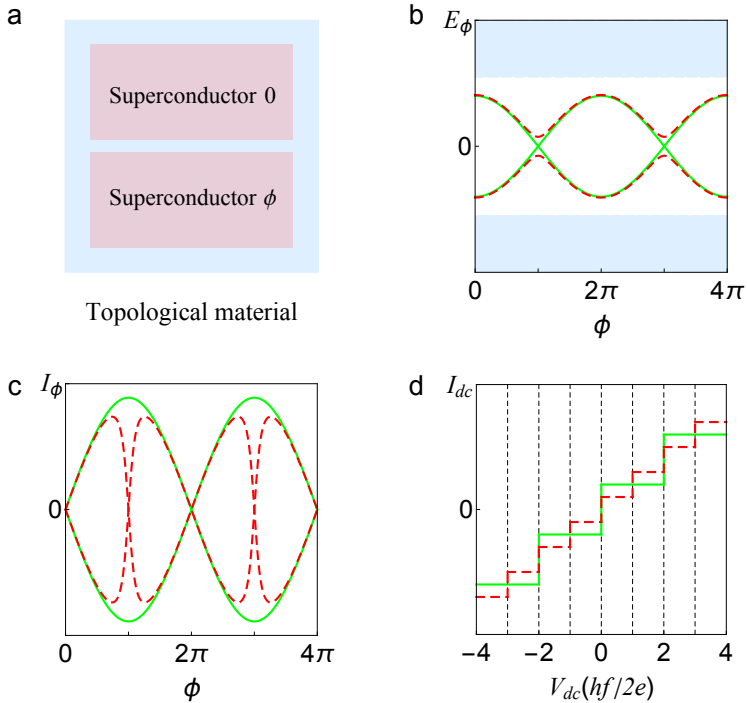

**Figure 1** The simplest fractional Josephson effect. **a**, Topological material proximity coupling to an *s*-wave Josephson junction placed on its surface. **b**, Andreev bound state spectra $E_\phi$ versus the phase difference $\phi$ across the junction in **a**; the crossings at $E_\phi = 0$ are Majoranas, and the blue shading represents the bulk continuum states of the superconducting regime. **c**, Current-phase relations $I_\phi$ based on the spectra in **b**. **d**, Shapiro steps in the dc current-voltage characteristic, when the junction in **a** experiences a dc bias of voltage $V_{dc}$ and a microwave irradiation of frequency *f*. In the presence of Majoranas at $E_\phi = 0$ in **b**, the current is *4π periodic* in **c**, and only even Shapiro steps are visible in **d**, as shown by the solid green curves. By contrast, in the absence of Majoranas, the current is *2π periodic* in **c**, and all integer Shapiro steps are visible in **d**, as shown by the dashed red curves. For simplicity, possible time-reversal symmetry, junction width, and dispersion along the junction are not taken into account here.

To achieve this result the researchers [2], by doping Bi with Sb, first fine-tuned their material to the transition point between topological and trivial band insulators. This produces an accidental bulk Dirac point neither protected by any symmetry nor connected to any surface Fermi arc. They used angle-resolved photoemission spectroscopy and magneto-transport to quantify the contributions to transport from unwanted surface states and the bulk Dirac cone, respectively. Next, they fabricated Josephson junctions of varying width and Nb-electrode separation on the surfaces of such accidental Dirac semimetals. Each junction shows a supercurrent as a clear manifestation of proximity-induced superconductivity, and moreover the $N = \pm 1$ Shapiro steps are missing for the shorter junctions. The researchers estimate that the $4\pi$ periodic contribution characteristic of Majoranas accounts for up to *20%* of the total critical current. In addition, they show that a small perpendicular orbital magnetic field decreases the total critical current whereas a small in-plane Zeeman magnetic field suppresses only the $4\pi$ periodic component.

The disappearance of odd Shapiro steps has been observed before in a Josephson effect mediated by helical edge states [10], but all experiments on the surfaces of topological insulators [11, 12] or Dirac semimetals [2, 13] only find the $N = \pm 1$ steps missing. This might be due to competing capacitive effects or relaxation processes. While the fractional Josephson effect in a topological Dirac semimetal $Cd_3As_2$ was attributed to surface-arc superconductivity [13], Li and co-authors consider the bulk Dirac cone responsible for the same effect in their accidental Dirac semimetal. Notably, 3D topological Josephson junctions can exist [14], and this fact might support the explanation by Li and co-authors. Going forward, it will be of paramount importance to solve these puzzles, to examine Josephson effects in other topological materials such as $(Bi_{1-x}In_x)_2Se_3$ [15] and 1T' $WTe_2$ [16], and to explore the interaction-driven $8\pi$ periodic Josephson effect that has been predicted as a signature of "fractional Majoranas" [7]. The missing Shapiro steps identified by Li *et al*. and other teams represent an important step in realizing topological superconductivity and fractional Josephson effects. Their results are truly desired and inspiring.

## Reference:

1. Josephson, D. *Phys. Lett.*, **1**, 251 (1962).

2. Li, C. *et al. Nat. Mater.* 17, 875-880 (2018).

3. Kitaev, A. *Phys. Usp.* **44**, 131 (2001).

4. Fu, L. & Kane, C. *Phys. Rev. Lett.* **100**, 096407 (2008).

5. Sau, J. *et al. Phys. Rev. Lett.* **104**, 040502 (2010).

6. Alicea, J. *et al. Nat. Phys.* **7**, 412-417 (2011).

7. Zhang, F. & Kane, C. *Phys. Rev. Lett.* **113**, 036401 (2014).

8. Shapiro, S. *Phys. Rev. Lett.* **11**, 80 (1963).

9. Jiang, L. *Phys. Rev. Lett.* **107**, 236401 (2011).

10. Bocquillon, E. *et al. Nat. Nanotech.* **12**, 137-143 (2017).

11. Rokhinson, L., Liu, X. & Furdyna, J. *Nature Phys.* **8**, 795-799 (2012).

12. Wiedenmann, J. *et al. Nat. Commun.* **7**, 10303 (2016).

13. Yu, W. *et al. Phys. Rev. Lett.* **120**, 177704 (2018).

14. Zhang, F. & Kane, C. *Phys. Rev. B* **90**, 020501(R) (2014).

15. Wu, L. *et al. Nat. Phys.* **9**, 410–414 (2013).

16. Wu, S. *et al., Science* **359**, 76–79 (2018).